\documentstyle[prl,twocolumn,aps,epsfig]{revtex}
\begin{document}
\draft
\wideabs{
\title{\bf{\LARGE{New way to achieve synchronization in spatially 
extended systems}}}
\author{Bikash C. Gupta$^{1}$, P. A. Sreeram$^2$ and S. B. Lee$^1$}
\address{$^1$Department of Physics, Kyungpook National University,
Taegu 702 701, Korea \\
$^2$ School of Physics and Astronomy,Tel Aviv University, Tel Aviv}
\maketitle
\begin{abstract}

We study the spatio-temporal behavior of simple coupled map lattices
with periodic boundary conditions. The local dynamics is governed 
by two maps, namely, the sine circle map and the logistic map respectively.
It is found that even though the spatial behavior is irregular for the
regularly coupled (nearest neighbor coupling) system, the spatially 
synchronized (sometime chaotically synchronized) as well as periodic 
solution may be obtained by the introduction of three long range couplings 
at the cost of three nearest neighbor couplings.
\end{abstract}
\narrowtext
}

Recently there have been lot of interests in the small world network 
systems \cite{watt1,watt2,watt3,amara,barra}.
The small world network systems lie in between the regular and the random
networks where only a small fraction of long range links are introduced at
the cost of the equal number of regular short range links keeping the
total number of links conserved. The properties of such systems have been
thoroughly studied very recently because of the fact that many biological, 
technological or social networks fall in this category. Intuitively it is 
expected that the information or source in the small world network will 
spread quickly through out the system because of the presence of some long 
range links.

On the other hand a number of phase models have been proposed over the
recent years to describe the dynamical behavior of large population of
nonlinear oscillators subject to a variety of coupling 
mechanisms \cite{choi1,Oopo1,walle,kane1}. A major
phenomenon that can be observed is the possibility of self synchronization
among the members of the population. These can represent the fire flies,
heart pace maker cells, pancreatic beta cells, neurons etc. as well as
the circuit arrays \cite{wies1}. Different cells (units) in different 
systems may be
locally governed by some specific rule. For example it may be governed 
by the sine circle map, logistic map or some differential equations
depending on the physical or biological system. Furthermore different
cells may be coupled to each other in various possible ways.   

Even if the local dynamics is regular, the spatial as well as the temporal 
behavior of the regularly coupled system may be irregular. It might happen
other way also, i.e., the dynamics of regularly coupled chaotic oscillators
may turn out to be regular. We are interested to confine ourselves in the 
parameter regime where the spatial behavior of a regularly coupled lattice 
becomes irregular. Then the question 
we ask is whether it is possible to control the spatial irregularity by
introducing some small fraction of long range couplings at the cost of the
equal number of nearest neighbor couplings keeping the total number
of couplings conserved. If so, then that would be a new way to control 
spatial irregularity in a spatially extended systems. Furthermore, if the
solution turns out to be synchronized spatially but chaotic temporally, 
it will find good application
to the secret communication network. We would like to investigate these aspects
in one dimensional lattice where the local dynamics of individual lattice
site is governed either by the sine circle map or the logistic 
map \cite{kane2}. The coupling we consider here is the simple unidirectional 
nearest neighbor coupling. First we discuss the unidirectionally coupled map 
lattice with the local dynamics defined by the sine circle map and then by 
the logistic map.

Here we consider a one dimensional periodic lattice of size $N$. The 
local dynamics at any site is governed by the following rule:
\begin{equation}
x\left(t+1\right)=f\left(x\left(t\right)\right)=x\left(t\right) 
+ \Omega -\frac{K}{2\pi} \sin\left(2\pi x\left(t\right)\right)
{\rm mod 1}
\end{equation}
where $x$ is the dynamical variable, $t$ is the time, $\Omega$ is the 
frequency and $K$ is the nonlinearity parameter. The map given by Eq. 
(1) is known as the  sine circle map and very much similar to the 
dynamical equation in the circuits of Josephson Junction arrays. If 
all the sites in the system are independent of each other, the dynamics
of the individual sites will be governed by Eq. (1). However, we are
interested in the dynamics of the system when all the sites are directly 
or indirectly coupled to each other. The lattice sites may interact 
with each other in a various possible ways. But, here, we consider the
situation where the dynamical variable at the $i^{th}$ site may be governed 
by the following rule:
\begin{eqnarray}
& &x_i(t+1)=(1-\epsilon) f(x_i(t)) + \epsilon f(x_{i+1}(t)) ; ~~1 \le i <N,
\nonumber \\
& &{\rm and } \nonumber \\  
& &x_N(t+1)=(1-\epsilon) f(x_N(t)) + \epsilon f(x_{1}(t))   
\end{eqnarray}
where $\epsilon$ is the coupling parameter. The spatio-temporal behavior 
of the system has been studied extensively. \cite{kane2} Even 
though the local dynamics 
is chaotic, the global dynamics of the system may turn out to be chaotic 
or regular depending on the parameters, initial condition, as well as 
the nature of the coupling. Our intention is to start with a random initial 
condition and fix the parameters such that the dynamics of the system 
becomes chaotic both spatially as well as temporally with the regular 
nearest neighbor coupling as described by Eq. (2). 

\vspace{.5cm}
     \begin{center}
      \epsfig{file=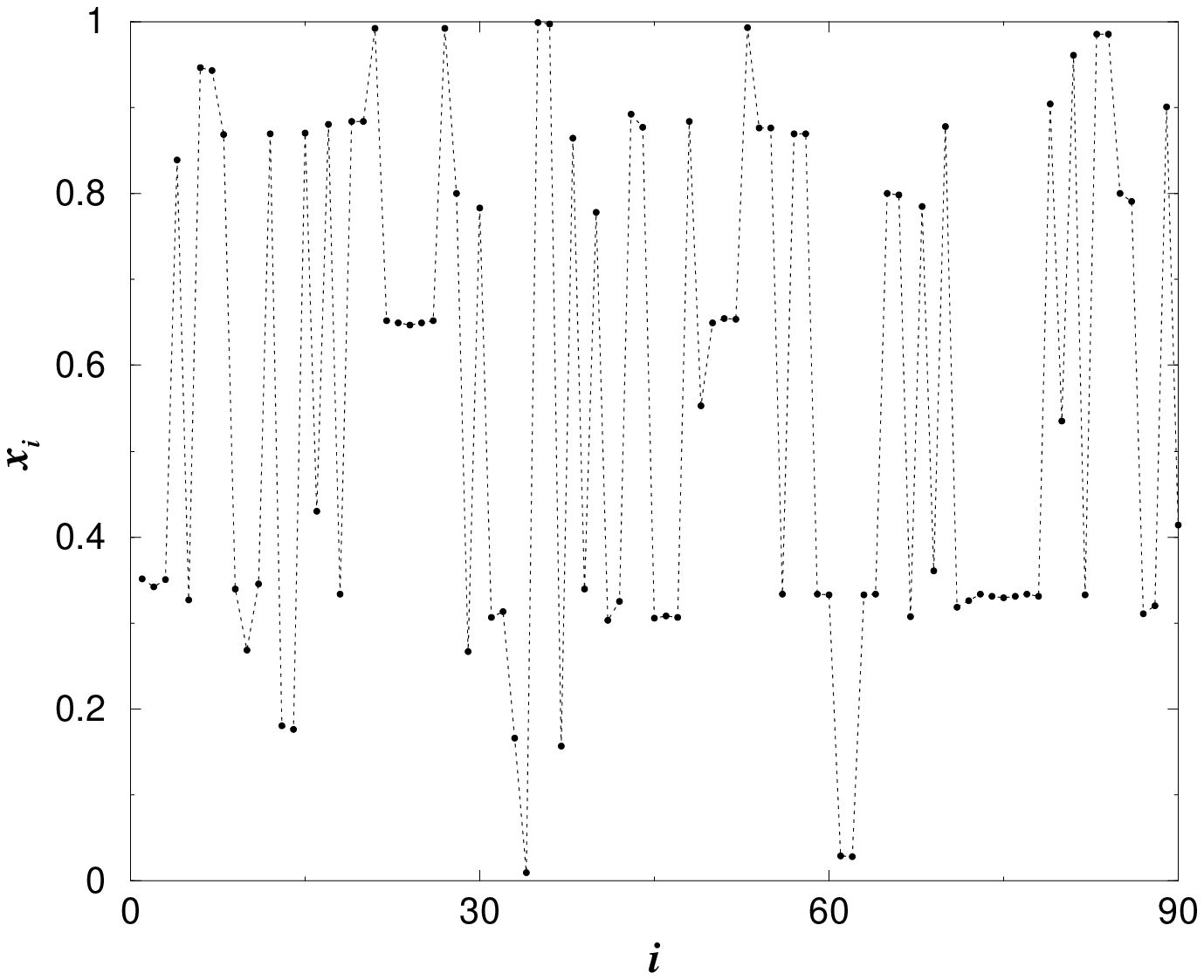,width=7.4cm,height=5.2cm}
      \vspace{0.5cm}
      \begin{figure}[p]
      \caption{All $x_i$'s at time $t$=5001 are plotted as a function of
               site index $i$ for a lattice of size 90 where all the sites
               are coupled to its nearest neighbor unidirectionally. The
               parameter values are $K=\sqrt 2$, $\Omega=0.3$ and $\epsilon=
               0.5$ respectively.
      \label{fig1}}
      \vspace{0.5cm}
      \end{figure}
      \end{center}

For example, we chose the parameter values as $\epsilon$=0.5, $\Omega$=
0.3 and $K=\sqrt{2}$. We consider the lattice size, $N$=90. We allow the
system to evolve for 5000 steps before determining it's state. The spatial
behavior of the system described by Eq. (2) is shown in Fig. (1) where 
$x_i$'s are plotted as a function of $i$ for t=5001. We see that there is no 
regular spatial behavior.  
The temporal evolution for all the sites are also found to be irregular.  
The temporal evolution at the first site is shown in the Fig. (2) as an
example. 

We introduce the concept of small world network systems in the spatially
extended system, namely, coupled map lattices. Suppose that we have a 
periodic one dimensional lattice of $N$ sites. If the lattice sites are 
coupled with its nearest neighbor site unidirectionally, there will be $N$
couplings or bonds. We will use the word bond in place of coupling for
convenience. A fraction of these $N$ bonds are replaced by long range 
couplings. The replacement of nearest neighbor bond (coupling) by the 
long range bond is made randomly. In the process of replacement, we 
make sure that duplicates are not allowed and no part of the system 
becomes isolated. Under such rearrangements, the total number of bonds 
remain conserved. Since the replacement of bonds 
are done randomly, there will be many possible configurations for fixed
number of bonds being replaced. Therefore, it is necessary to check if 
any of those configurations produces the synchronized solutions.   

\vspace{.5cm}
     \begin{center}
      \epsfig{file=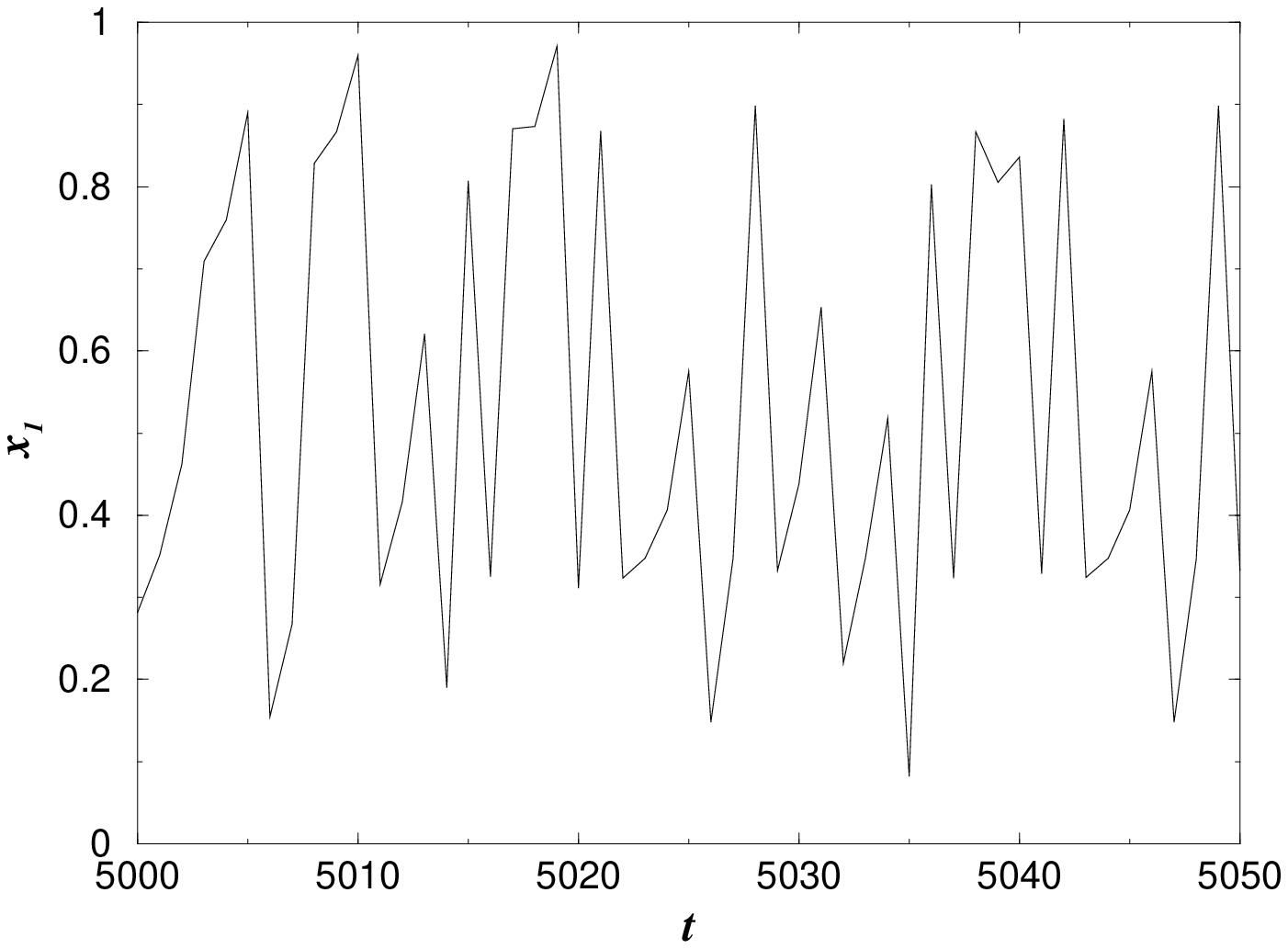,width=7.4cm,height=5.2cm}
      \vspace{0.5cm}
      \begin{figure}[p]
      \caption{This shows the temporal evolution of the dynamical variable
               $x$ at site 1 for the same system with same parameters as
               used in Fig. 1.
      \label{fig2}}
      \vspace{0.5cm}
      \end{figure}
      \end{center}
So, we consider the system  where $p$ 
fraction of nearest neighbor bonds are replaced by randomly chosen 
long range bonds.  The parameter values are are chosen to be same as 
in Fig. (1).  Initially all the sites are populated randomly and then 
they evolve according to Eq. (2) with the replacement of $p$ fraction 
of bonds.  The solution appears to be stable after 5000 iterations. 
Since there are large number of possible configurations, the numerical 
experiment is carried out for 1000 configurations for a fixed value of 
$p$. The simulation is also carried out for different values of $p$ 
between 0 and 1. It is found that only two to ten percentage of configurations 
lead to synchronized solutions for each value of $p$. For $p=0.5$ more 
number of configurations produce synchronized solution. 

Thus we experience that there exist some configurations which may produce 
the synchronized solutions. Therefore, the control of the spatially 
chaotic behavior is possible by the above mentioned mechanism but the 
percentage value of configurations producing the synchronized solutions 
are very low. So, one has to go through various sets of
re-wiring to achieve the synchronization and is therefore difficult to 
implement in the physical systems. One needs to know a re-wiring procedure 
where least number of bonds will be disturbed or replaced. Furthermore, one 
should know a systematic way of replacement of bonds to achieve the 
synchronization. 
Intuitively we suspect that the replacing of the nearest neighbor bonds
may be made in a systematic manner and think that the bonds should be
broken at regular interval in the lattice and the disturbed vertices 
should be connected in a regular fashion. More explicitly, let us
divide the lattice into $n$ number of segments, each segment has length 
(L) equal to $\frac{N}{n}$. So, the bond between sites $i$ and $i+1$ should 
be broken and the $i^{th}$ site should be connected to $(i+L)^{th}$ site. 
Next, the bond between sites $(i+L)$ and $(i+L+1)$ should be broken and 
$(i+L)^{th}$ site should be connected to $(i+2L)^{th}$ site and so on. 
This procedure should be continued until the full lattice is covered. 
We note that if $n=N$ or $L=1$, then the modified lattice 
remains same as the regular one. Therefore, we expect that $n$ should be 
between 1 to $N-1$. 
We therefore do the numerical experiment for $n$=1, 2, 3, 4 and so on. In fact 
we find that $n$=1 and 3 cases work very well. Although, ideally, $n$=1, 
would be the most suitable solution, we also study $n$=3, since it is the
most ideal case for studying multiple rewirings. We find that $n$=2 and $n$=4
have lesser probability to synchronize than $n$=3, over wide regions of the
$\Omega-\epsilon$ phase space. We will discuss these concepts in details in
a forthcoming article. For $n$=3, the re-wiring rule is given explicitly for 
a lattice of size 90: 
\vspace{.5cm}
     \begin{center}
      \epsfig{file=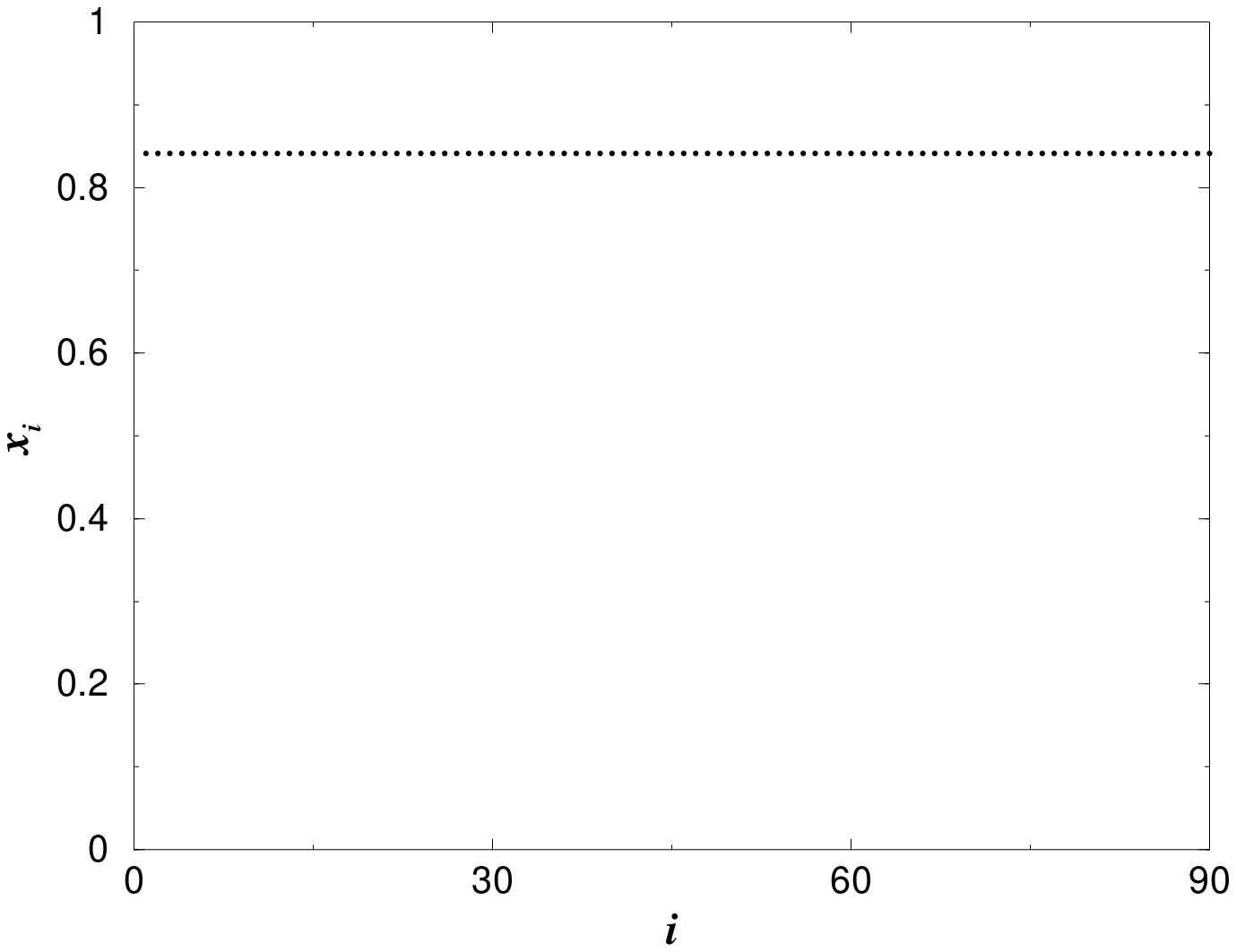,width=7.4cm,height=5.2cm}
      \vspace{0.5cm}
      \begin{figure}[p]
      \caption{All $x_i$'s at time $t=5001$ are plotted as a function of
               site index $i$ for a system of size 90 where three long
               range couplings are introduced at the cost of three 
               nearest neighbor couplings. The parameter values are same 
               as used in Fig.1  
      \label{fig3}}
      \vspace{0.5cm}
      \end{figure}
      \end{center}
The bonds between sites 1 and 2, 30 and 31, and 60 and 61 are broken and 
the links between sites 1 and 30, 30 and 60, 60 and 90 are established.
Furthermore, any arbitrary site can be chosen as site 1. The spatial 
behavior for this configuration is shown in Fig. (3). The parameters
values are same as used for Fig. (1). We notice that total synchronization
is achieved. The synchronization behavior can be obtained for a lattice 
of any size with the above mentioned rule. The only difference is that
for larger lattice it takes longer time to achieve the synchronization.
This is in fact not surprising. However, the temporal evolution is
chaotic. So, it may be called chaotic synchronization. We further note
that even if the site 1 is coupled with any site around the $30 th$ site
with small width, synchronization is achieved. Thus a small defect in
re-wiring does not matter in achieving the synchronization. 

We clearly note that by disturbing only three regular nearest neighbor bonds 
and establish three long ranged bonds in a regular way, one can achieve
synchronization over a wide range of parameters.
The phase diagram in the $\epsilon - \Omega$ plane for $k=\sqrt{2}$
is shown in Fig 4. The shaded region (in fig. 4) in the $\epsilon - \Omega$ 
plane are the allowed parameter regime where one achieves the synchronized 
solution starting from a random initial conditions. We see that there
is a large region in the $\epsilon-\Omega$ plane where the synchronization
is obtained through the re-wiring mechanism. The synchronized region for
only one broken bond is larger than that of the 3 bond case by about
20-30 $\%$. However, for any other number of rewirings, the allowed part of 
the phase space is much smaller.

\vspace{.5cm}
     \begin{center}
      \epsfig{file=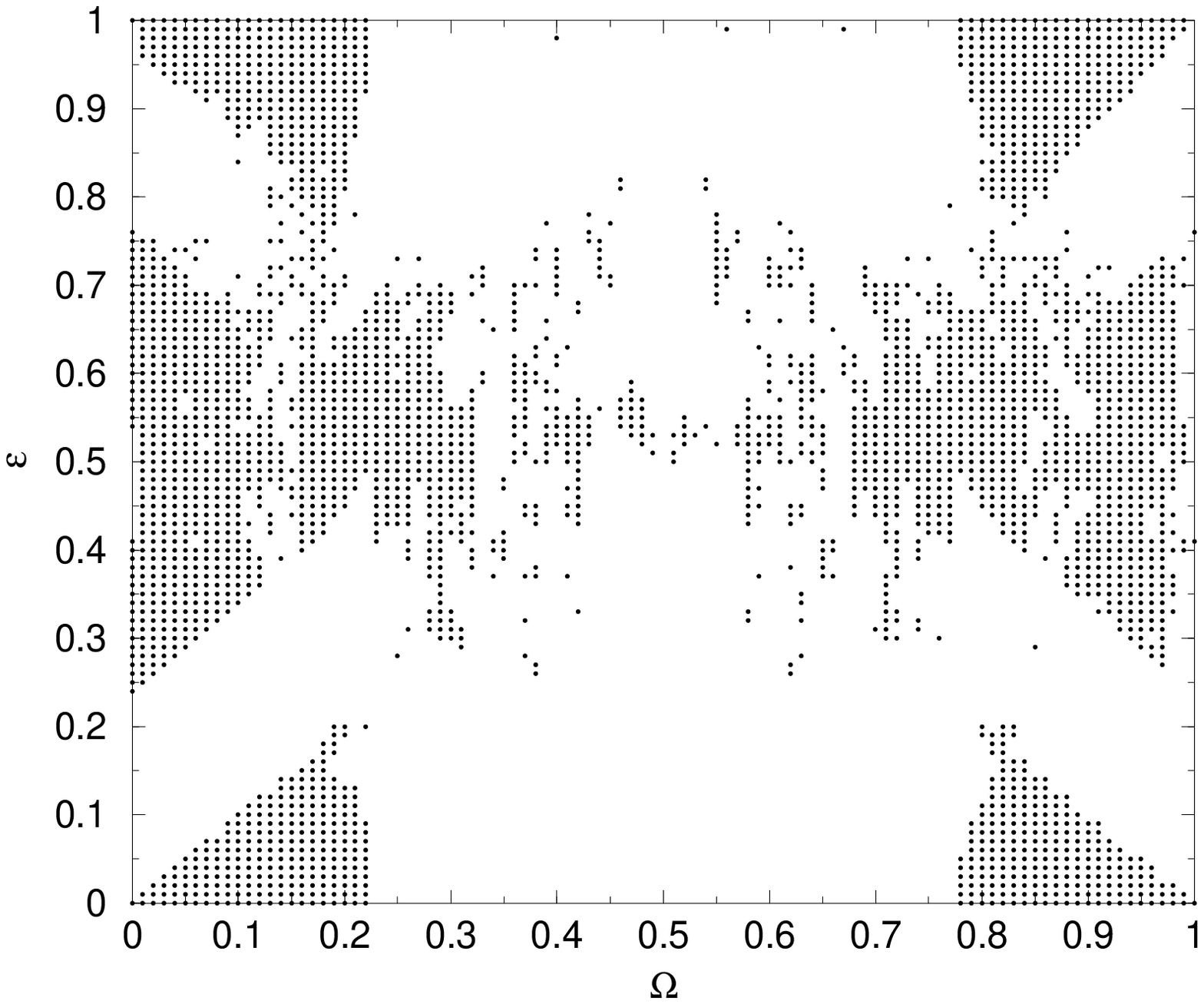,width=7.4cm,height=5.2cm}
      \vspace{0.5cm}
      \begin{figure}[p]
      \caption{Phase diagram in the $\Omega-\epsilon$ plane for the same 
               system as described in Fig. 3. Here $K=\sqrt 2$. The shaded
               region corresponds to the allowed parameter space to obtain
               the chaotic synchronization with the re-wiring mechanism.
      \label{fig4}}
      \vspace{0.5cm}
      \end{figure}
      \end{center}

Here we consider  the same system where the local dynamics is governed 
by the Logistic map. The logistic map has been extensively used to model 
a number of physical as well as biological phenomenon \cite{kane2}. 
The map is given by
\begin{equation}
x(t+1)=f\left(x(t)\right)=\mu x(t) \left( 1 - x(t)  \right)
\end{equation}
where $\mu$ is the parameter controling the local dynamics.
The the dynamics of the variable $x_i$ at the $i^{th}$ site is governed 
by the Eq. (2) with $f\left(x(t)\right)$ as given in Eq. (3)
where $t$ represents the time, and $\epsilon$ is the coupling strength.
In this system also we find that one may obtain the spatially periodic
solution and sometimes synchronized solution through the re-wiring rule 
as stated earlier. 

In this system, we find that there is a critical value of the coupling
strength $\epsilon$ for each value of $\mu$. For example, all
$x_i$'s are plotted as a function of $\epsilon$ for $\mu$=4 in Figs. (5) and 
(6). Fig. 5 is for the system where all the sites are regularly coupled
to its nearest neighbor. But Fig. 6 is for the rewired lattice. 
We clearly see from Fig. 5 that there is no spatial regularity.
On the other hand, for the rewired lattice (Fig. 6) there is a critical 
value of $\epsilon$, say $\epsilon_{cr}=0.35$ above which the solution is 
spatially periodic of period four.  The spatial nature of the solution 
for the rewired lattice is verified in the entire $\epsilon-\mu$ plane 
as shown in Fig 7. There are various regions 
corresponding to various kinds of solutions. The solution is spatially
periodic with period one or four in the region I, the solution is 
temporally as well as spatially irregular in region II, the solution is
spatially periodic of period four in region III and the solution is 
chaotically synchronized in region IV.

\vspace{.5cm}
     \begin{center}
      \epsfig{file=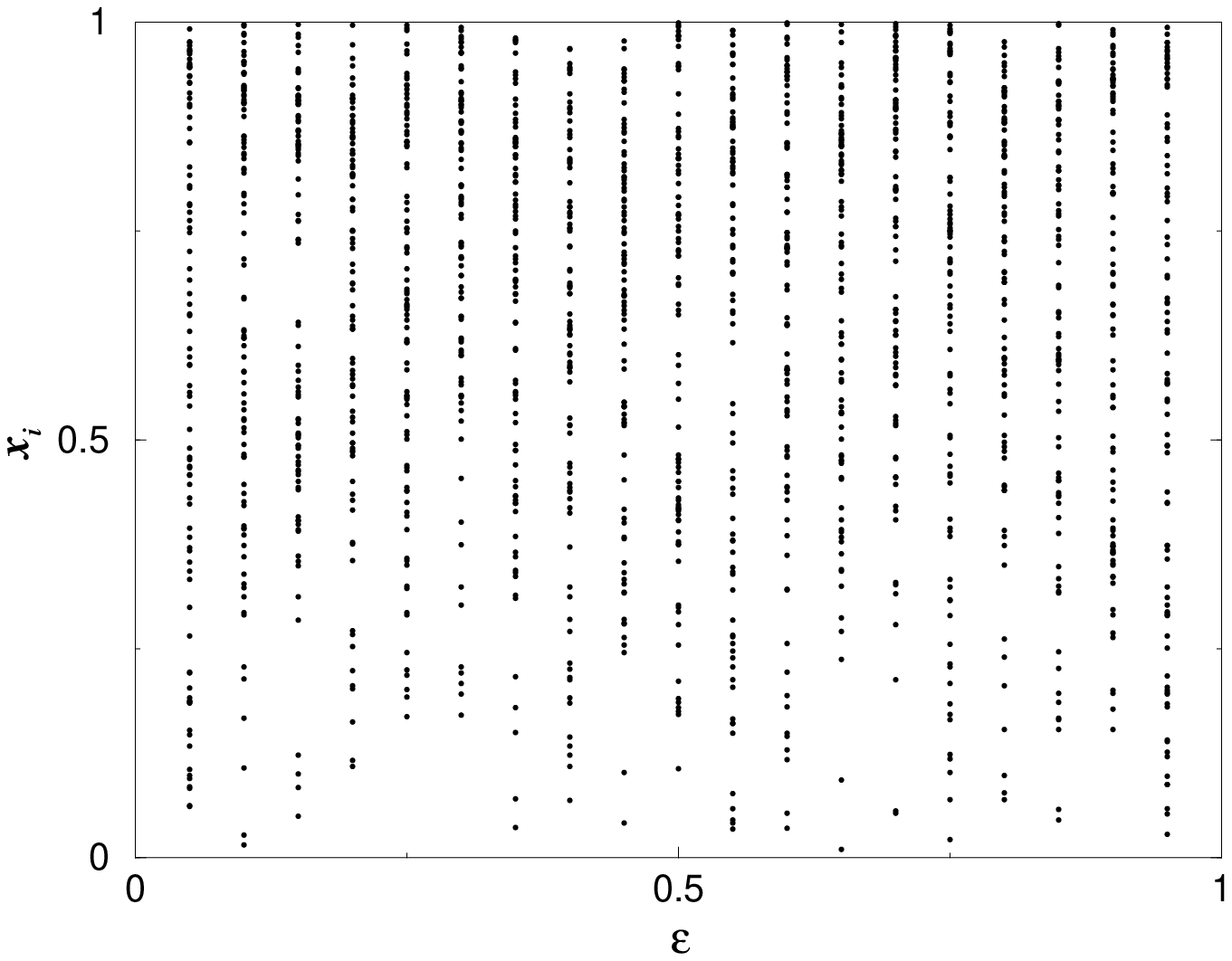,width=7.4cm,height=5.2cm}
      \vspace{0.5cm}
      \begin{figure}[p]
      \caption{All $x_i$'s at time $t=5001$ are plotted as a function of 
               $\epsilon$ for a system of size 90 where all the lattice
               sites are coupled to its nearest neighbor unidirectionally.
               Here $\mu=4$.
      \label{fig5}}
      \vspace{0.5cm}
      \end{figure}
      \end{center}
\vspace{.5cm}
     \begin{center}
      \epsfig{file=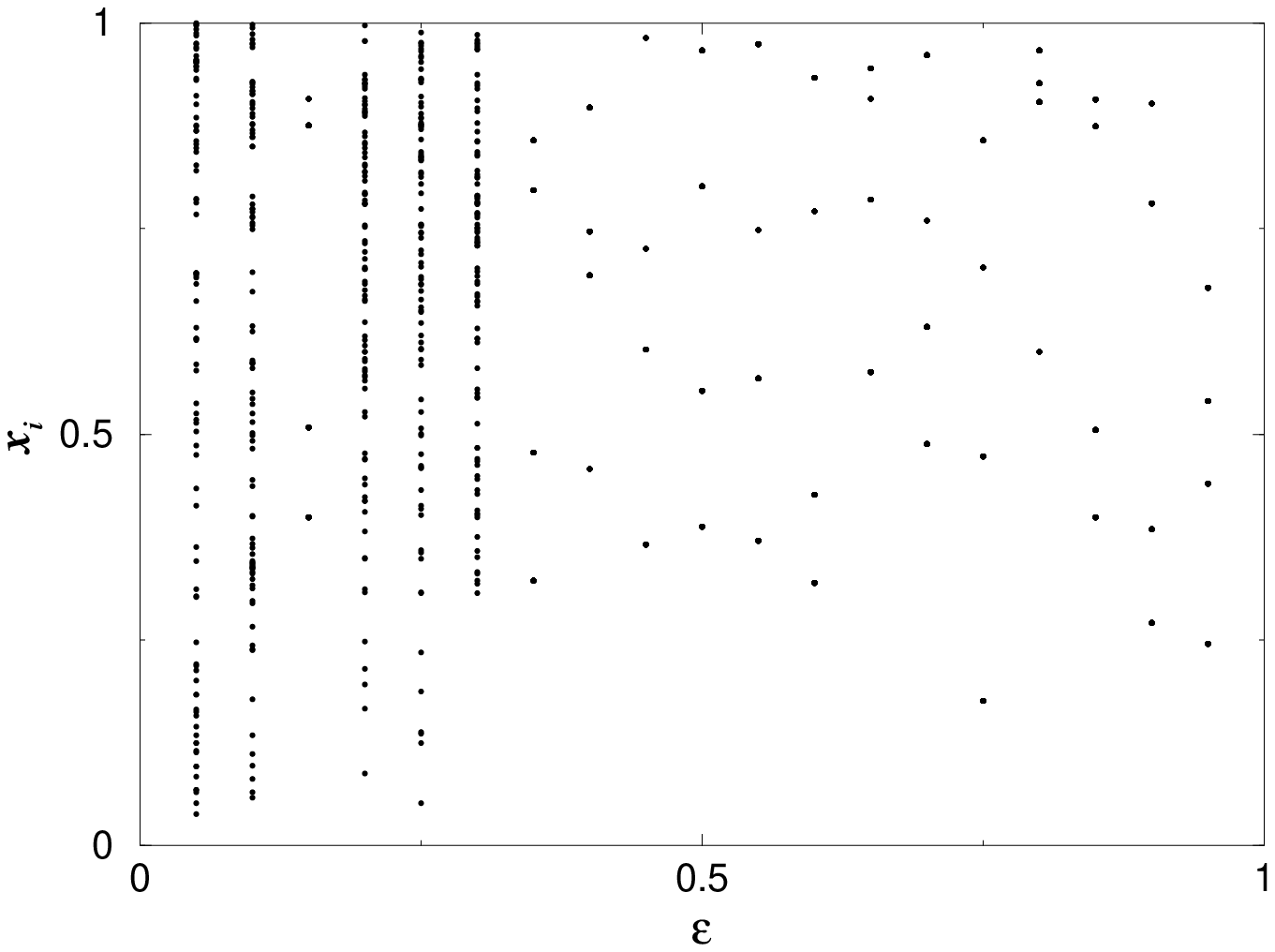,width=7.4cm,height=5.2cm}
      \vspace{0.5cm}
      \begin{figure}[p]
      \caption{All $x_i$'s at time $t=5001$ are plotted as a function 
               $\epsilon$ for a system of size 90 where three long ranged
               couplings are introduced at the cost of three nearest
               neighbor couplings. Here $\mu=4$.
      \label{fig6}}
      \vspace{0.5cm}
      \end{figure}
      \end{center}
Thus we see that the re-wiring mechanism helps us to control the spatial 
irregularity in a large region of the parameter space. Furthermore we 
note the parameter region for both the spatial 
as well as temporal periodic solution also large. Therefore, this 
mechanism with the logistic map works well to control spatial and temporal
irregularity in a large region of parameter space.

\vspace{.5cm}
     \begin{center}
      \epsfig{file=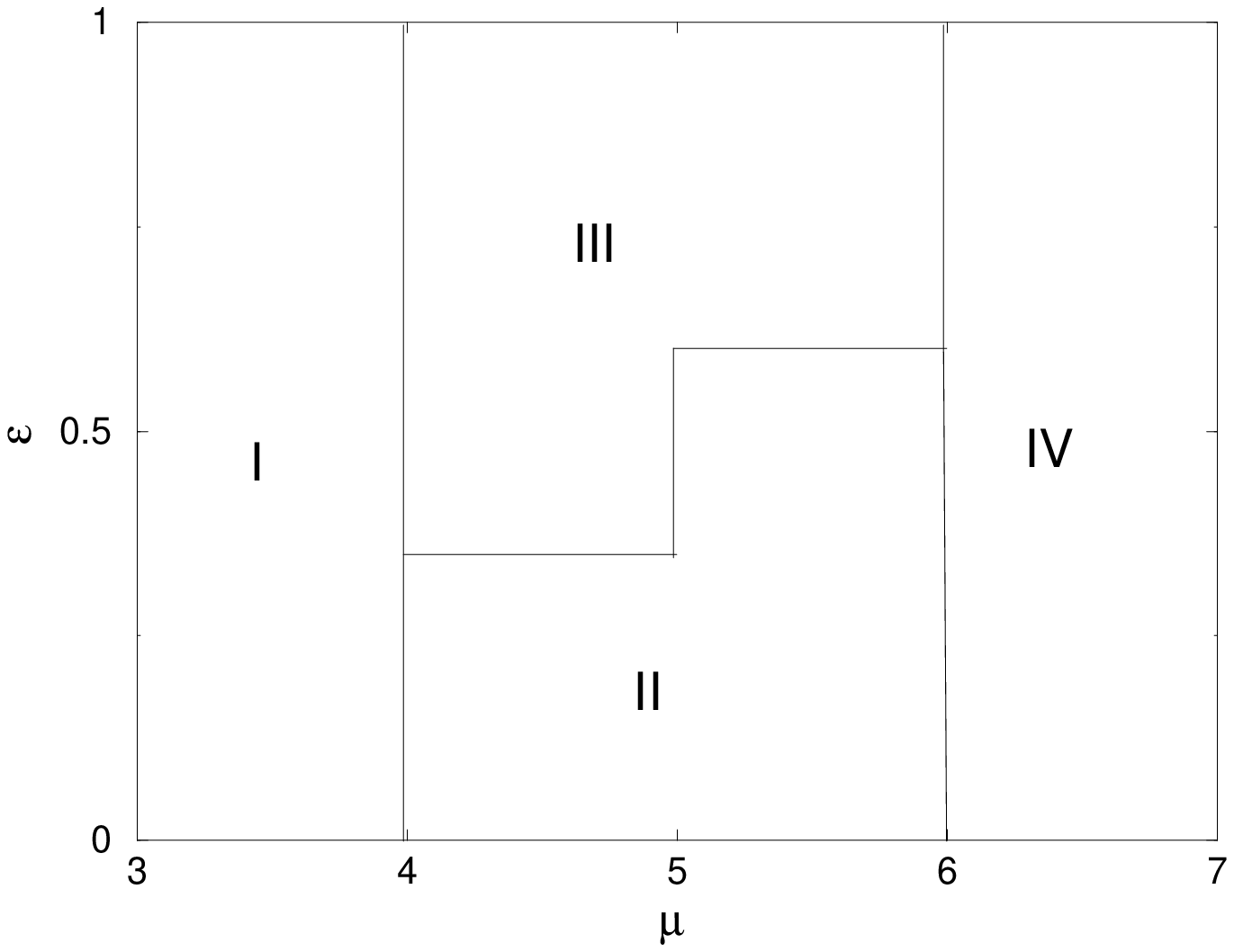,width=7.4cm,height=5.2cm}
      \vspace{0.5cm}
      \begin{figure}[p]
      \caption{Phase diagram in the $\epsilon-\mu$ plane for the system as
               described in Fig. 6. There are four
               regions corresponding to four kinds of solutions. The solution
               is periodic of period one or four in region I, spatially and 
               temporally irregular in region II, spatially periodic of 
               period four in region III and chaotically synchronized in 
               region IV. 
      \label{fig7}}
      \vspace{0.5cm}
      \end{figure}
      \end{center}

In summary, we have studied spatially extended system where the local 
dynamics are 
governed by the sine circle map and the logistic map. We have implemented 
the idea of small world network and investigated that spatial irregularity
in the system can be controlled. We have shown that for a regularly coupled 
system showing irregular behavior spatially, synchronization can be achieved
by disturbing a few regular links (nearest neighbor coupling) and establishing
an equal number of long ranged links in a regular way. We notice that the
region in the $\Omega-\epsilon$ space where synchronization occurs depends
on how many of these rewirings are done. For one and three rewirings, 
synchronization occurs over large regions of the $\Omega-\epsilon$ space.
We will establish the reason for this in a forthcoming article.
Chaotic synchronization is achieved for the sine circle
map with the re-wiring mechanism. On the other hand, spatially as well as 
temporally periodic solution is achieved for the
logistic map. Thus we experience that the re-wiring mechanism helps us
to achieve the chaotic synchronization for some local dynamics. The chaotic
synchronization is an important solution in Josephson junction arrays. 
Furthermore it is very useful as far as the secret communication network is 
concerned. This mechanism may be tested with other local dynamics existing
in physical or biological systems.

\end{document}